\documentclass{article}
\usepackage{graphicx} 
\usepackage{color, soul}
\usepackage{cite}
\usepackage{subcaption}
\usepackage{ulem}
\usepackage{authblk}
\usepackage{hyperref}

\title{Neutron stars in $f(Q) = Q +\xi Q^2$ gravity}

\author[1]{J.C.N. de Araujo\thanks{jcarlos.dearaujo@inpe.br}}
\author[1,2]{H.G.M. Fortes\thanks{hemily.gomes@gmail.com}}
\affil[1]{Divisão de Astrof\'{i}sica, Instituto Nacional de Pesquisas Espaciais, Avenida dos Astronautas 1758, S\~{a}o Jos\'{e} dos Campos, SP 12227-010, Brazil}
\affil[2]{Departamento de Física de Volta Redonda, Universidade Federal Fluminense, Campus Aterrado, Rua Desembargador Ellis Hermydio Figueira 783, Volta Redonda, RJ, 27213-145, Brazil.}

\date{}

\begin{document}

\maketitle

\begin{abstract}

{Modified theories of gravity based on the symmetric teleparallel framework have recently attracted considerable attention as viable alternatives to General Relativity. In this context, $f(Q)$ gravity, in which the gravitational interaction is encoded in the nonmetricity scalar $Q$, provides a consistent geometrical formulation that differs from the standard curvature-based description. In this work, we investigate the structure of neutron stars within a family of $f(Q)$ gravity models using four realistic equations of state: FPS, SLy, ENG and MPA1.
Using the covariant formulation of $f(Q)$ gravity, we derive the corresponding Tolman–Oppenheimer–Volkoff equations and apply them to model compact stellar configurations. Numerical integration of the field equations provides the mass–radius relations and the maximum masses supported by each equation of state, enabling a direct comparison with current observational constraints. Furthermore, we analyze the behavior of the nonmetricity scalar both inside and outside the stellar object, providing additional insight into the gravitational structure of compact stars in this framework.}
\end{abstract}

\section{Introduction}
\label{int}

General relativity (GR) has been remarkably successful in describing gravitational phenomena on a wide range of scales, from laboratory experiments to astrophysical and cosmological observations \cite{Will2}. Nevertheless, the increasing amount of high-precision cosmological data has motivated the investigation of modified theories of gravity as possible extensions of Einstein’s theory, particularly in view of open problems such as the late-time accelerated expansion of the Universe and the nature of dark energy \cite{Riess,Peebles,Capozziello2}. 

{In the standard formulation, the gravitational interaction is based on the curvature of spacetime $R$, the connection is metric-compatible ($\nabla_\alpha g_{\mu\nu}=0$) and torsionless (symmetric), which leads to the standard formulation of GR. As an extension of GR, we can cite the $f(R)$ models \cite{fR} which offer a framework in which the Einstein-Hilbert action is replaced by a more general action involving an arbitrary function of the scalar curvature $R$. Such models have been extensively studied in recent decades. However, the field equations in these models are typically fourth-order differential equations, which complicates their analysis.}

{The curvature-based approach is not the only way to describe gravity. The so-called Geometrical Trinity of Gravity \cite{Jimenez2} provides equivalent formulations based on geometrical objects beyond curvature.}

{The first formulation is Teleparallel Theory of Gravity \cite{Aldro,review2,TT}, which is known to be equivalent to GR, with the fundamental difference that gravity is described by the torsion scalar $T$ rather than the curvature scalar $R$. In analogy with $f(R)$ gravity, the natural extension of Teleparallel Gravity leads to $f(T)$ theories \cite{Ferraro,Linder}. These models have the advantage of producing second-order field equations, which are therefore simpler than the fourth-order equations typically obtained in $f(R)$ gravity. Moreover, $f(T)$ gravity has provided a variety of interesting cosmological and astrophysical solutions, offering alternative interpretations of the accelerated expansion of the Universe \cite{Ferraro,Linder,Myr,Karami} and allowing applications to compact stellar  objects \cite{Ganiou,Kpa,Pace,Ilijic,Bohmer,Pace2,FA1,FA2,FA3,FA4,FA5}.
Although GR and Teleparallel Gravity are equivalent at the fundamental level, this correspondence is generally lost in their modified versions, such as $f(R)$ and $f(T)$ gravity. 
For this reason, it is important to not overlook the possibility that such alternative models may lead to physically relevant results.}

{While TEGR and its extensions, such as $f(T)$ gravity, differ from GR due to the presence of a non-vanishing antisymmetric part of the connection, namely torsion $T$, both $TEGR$ and $GR$ preserve metricity, i.e., $\nabla_\alpha g_{\mu\nu}=0$. This motivates a third geometrical description of gravity, known as the Symmetric Teleparallel Equivalent of General Relativity (STEGR), in which metricity is no longer preserved and gravity is instead described by the nonmetricity tensor which is defined as $Q_{\alpha\mu\nu} \equiv \nabla_\alpha g_{\mu\nu}$. In this framework, the gravitational action is constructed solely from nonmetricity and reproduces the dynamics of GR {\cite{Jimenez2,Heisenberg}}.}

{The Symmetric Teleparallel Equivalent of General Relativity can be generalized by promoting the nonmetricity scalar in the action to an arbitrary function, leading to the so-called $f(Q)$ gravity. 
{Applications of $f(Q)$ gravity have been investigated in cosmology \cite{Mandal,Ferreira,Shekh,Jimenez}, in compact-star and gravastar models \cite{Maurya,Lohakare,Errehymy,Lin,Sokoliuk,Bhar,Pradhan2,FA6,Alwan,Heisen,Dimakis}, and in static, spherically symmetric configurations, including black-hole and wormhole solutions \cite{Calza,Wang}. The formal structure, covariant formulation, and viability of the theory have also been studied \cite{Ambrosio,Zhao,De}, see also the review in \cite{Heisenberg}. Additional studies have explored the cosmological, astrophysical, and theoretical aspects of $f(Q)$ gravity and related theories within the symmetric teleparallel framework \cite{Pradhan,Bhardwaj,Myrzakulov,Adak,Zhang,Ambrosio,Blixt,Ferreira,Maurya,Lohakare,Shekh,Capozziello,Gakis,Zhao,Wang,De,Pradhan2,Gadbail2}.}

In a previous study, we investigated the same class of $f(Q)$ gravity models using polytropic equations of state (EOSs) \cite{FA6}, which provided initial insights into the behavior of compact objects in these theories. In the present work, we study $f(Q)$ gravity models employing realistic equations of state, namely, FPS, SLy, ENG, and MPA1, and analyze their implications for the structure of compact stars.}

{Within GR, the compactness of a static spherically symmetric star is bounded by the Buchdahl limit \cite{Dadhich,buch1959}, which imposes the condition $M/R \leq 4/9$. This theoretical bound raises the question of whether alternative descriptions of gravity may allow configurations with compactness higher than those predicted in GR. In modified gravity scenarios, the stellar equilibrium and stability conditions can be altered, potentially affecting the maximum mass supported for a given equation of state (see, e.g., \cite{Nunes}). Such deviations are particularly relevant in light of recent observations, since they may provide a framework to interpret unusually massive compact objects, such as the secondary component of the GW190814 event \cite{GW} {of 2.6 M$_\odot$}, as a possible high-mass neutron star.}

In the context of modified gravity, several studies have investigated the properties of compact objects within specific alternative models { \cite{FA1,FA2,FA3,FA4,FA5,FA6,Alwan,Heisen,Dimakis}, using different models and approaches and focusing on how deviations from GR may affect stellar structure, maximum masses, and compactness. In particular, the paper \cite{Alwan} studied quadratic, exponential and logarithmic models with realistic equations of state. For the quadratic model, they reported that their analysis did not yield larger masses than in GR. Using a different approach, \cite{Dimakis} studied relativistic stars for $f(Q)$ gravity, obtaining solutions for a specific choice of spacetime and analyzing the implications for the properties of the star. These works are examples of  methods currently applied to compact stars in $f(Q)$ gravity and they indicate that certain functional forms of $f(T)$ and $f(Q)$ gravity can accommodate limits less restrictive than those typically found in GR. The present work focuses on the quadratic model $f(Q)=Q+\xi Q^2$ and extends our previous analysis with polytropic equations of state \cite{FA6} to four realistic EOSs, namely, FPS, SLy, ENG and MPA1. These realistic EOSs provide a more accurate description of dense matter in compact stars. This allows us to investigate how the quadratic correction modifies gravity and, in turn, affects stellar structure, maximum masses, and compactness under more astrophysically realistic conditions. We also examine whether the results obtained in simplified models persist when realistic microphysics is taken into account. We also compare the gravitational mass $M_S$ with the total rest mass $M_0$ to identify configurations that satisfy the condition $M_S<M_0$. The resulting sequences are confronted with the observational constraints shown in the mass-radius figures.}

{This paper is organized as follows. In Section~\ref{be}, we present the basic field equations of $f(Q)$ gravity and derive the modified TOV equations. In Section~\ref{sectionMS}, we specify the functional form of $f(Q)$ and present the differential equation system to be solved. In Section~\ref{Ne}, we consider realistic equations of state and perform the numerical analysis to obtain the mass-radius and mass-central density sequences for neutron stars {and compare our mass-radius relations with observational data for selected neutron stars.}
We also discuss the nonmetricity and the behavior of metric functions. Finally, in Section~\ref{Fr}, we summarize our conclusions and discuss possible future developments.}

\section{{The Tolman-Oppenheimer-Volkoff equations}
\label{be}}

{The basic equations of the extended Symmetric
Teleparallel Equivalent of General Relativit have been presented in several works in the literature. Here, we refer the reader in particular to studies adopting the covariant formulation of $f(Q)$ gravity, such as \cite{Zhao,Blixt}, since this is the framework employed here. This choice is essential because, when the coincident gauge is imposed at the level of action, where the affine connection is set to vanish, this may, in some cases, restrict the theory to solutions that reproduce GR. The covariant formulation overcomes this limitation by allowing for a non-vanishing connection, making it possible to consistently construct non-linear extensions in $Q$. In this framework, the general equations of motion can be derived without restricting the geometric structure from the outset. The covariant approach has been widely applied in the literature in different contexts, {such as cosmological scenarios \cite{Heisenberg,Jimenez} and compact-star configurations, including anisotropic stars, hybrid stars, and neutron stars \cite{Errehymy,Sokoliuk,Bhar,Alwan}.}

{A general review of $f(Q)$ gravity can be found in \cite{Heisenberg} while our previous work \cite{FA6} provides a brief discussion of the framework and its application to compact stars.} Here, we present only the main equations. The starting point is the definition of the nonmetricity scalar, given by
\begin{equation}
    Q = - P^{\alpha\beta\gamma} Q_{\alpha\beta\gamma},
\end{equation}
where $Q_{\alpha\beta\gamma}$ is the nonmetricity tensor, defined as
\begin{equation}
    Q_{\alpha\beta\gamma} = \nabla_\alpha g_{\beta\gamma}.
\end{equation}
The tensor $P^{\alpha\beta\gamma}$ is the nonmetricity conjugate, expressed in terms of the nonmetricity tensor and its contractions (see, e.g., \cite{FA6,Heisenberg,Zhao}).}

{As usual, the field equations are obtained from the action
\begin{equation}
S = \int \left( \frac{f(Q)}{16\pi} + \mathcal{L}_m \right) \sqrt{-g}\, d^4 x ,
\end{equation}
{The covariant action and its variation with respect to the metric are discussed in detail in Refs. \cite{Heisenberg,Zhao,Jimenez}.}

{Varying this action with respect to the metric leads to the field equations
\begin{equation}
2{P^\alpha}_{\mu\nu}\frac{d^2f}{dQ^2}\partial_\alpha Q
+\frac{1}{2}g_{\mu\nu} \left(f - Q\frac{df}{dQ} \right)
+ \frac{df}{dQ} G_{\mu\nu}
= 8\pi T_{\mu\nu}.
\label{eom1}
\end{equation}
For spherically symmetric stars, the appropriate metric is
\begin{equation}
    ds^2=e^{A(r)} dt^2-e^{B(r)} dr^2-r^2 d\theta^2-r^2 \sin ^2 \theta \, d\phi^2.
    \label{metric}
\end{equation}
The nonmetricity scalar then takes the form
\begin{equation}
  Q(r) = \frac{\left(e^{-B}-1 \right)\left( A' + B' \right)} {r},
  \label{Qr}
\end{equation}
where the prime denotes differentiation with respect to the radial coordinate $r$.}

{Since we are dealing with realistic equations of state, the energy–momentum tensor is taken to be that of a perfect fluid. Consequently, from the field equations, we obtain the following set of independent equations:
\begin{equation}
16\pi r^2\rho e^B = 2rf'_Q(e^B-1)+f_Q[(e^B-1)(2+rA')+(e^B+1)rB']+fr^2e^B ,
\label{E00}
\end{equation}
\begin{equation}
16\pi r^2 P e^B = -2rf'_Q(e^B-1)-f_Q[(e^B-1)(2+rA'+rB')-2rA']-fr^2e^B ,
\label{E11}
\end{equation}
\begin{equation}
32\pi r P e^B = 2rf'_QA'-f_Q[2A'(e^B-2)-rA'^2+B'(2e^B+rA')-2rA'']-2fre^B ,
\label{E22}
\end{equation}
where $\rho$ and $P$ denote energy density and pressure, respectively.}

{Adding Eqs.~(\ref{E00}) and (\ref{E11}) leads to a useful relation
\begin{equation}
(A' + B')f_Q =8\pi r(\rho+P)e^B .
\label{AlBlFQ}
\end{equation}
In vacuum, this equation implies $A' + B' = 0$, which coincides with the result obtained in GR. Consequently, the vacuum solutions in $f(Q)$ gravity reduce to those of GR.}

\section{Neutron stars in $f(Q) = Q +\xi Q^2$ gravity}
\label{sectionMS}

{In a recent publication \cite{FA6}, we considered in detail the set of equations required to model polytropic stars within a specific \( f(Q) \) gravity model, namely,
\begin{eqnarray}
f(Q)= Q + \xi \, Q^2 \, ,
\label{fQ}
\end{eqnarray}
where \( \xi \) is a real constant. Here, we consider realistic EOSs instead of polytropic ones. However, the differential system to be solved remains the same. Therefore, we present only the equations required in the present analysis and refer the reader to \cite{FA6} for a complete derivation.}

{In short, to model neutron stars for a given EOS, one must solve a system of differential equations for the metric functions \( A(r) \) and \( B(r) \), obtained from the field equations presented in the previous sections, together with an additional differential equation obtained from the conservation equations, which relates \( A'(r) \) and \( P'(r) \).}

{For a given EOS, the system of differential equations to be solved is:}
{
\begin{eqnarray}
    A'' &=& \biggl\{ A'\xi r(e^B-1)(A'+B')[4A'e^B-(A'+B')(3e^B+1)] + \nonumber \\
    & & -2\xi(e^B-1)^2 (A'+B')[(A'+B')(e^B+5)-2A'] + \nonumber \\
    & & + A'r^2 e^B[(A'+B')(e^B+1)-2A' e^B] + re^B(e^B-1)(4A' + 6B') + \nonumber \\
    & & -16\pi r^2 \rho \,e^{2B}[2(e^B-1) + A'r ]\biggr\}\cdot \nonumber \\
    & & \cdot
        \biggl\{ 2r(e^B - 1)[2\xi(A'+B')(1-e^B) + r e^B]\biggr\}^{-1},
        \label{All}
\end{eqnarray}}

{
\begin{eqnarray}
A' + B' = \frac{r \, e^B}{4\xi(e^B-1)}\left[ 1 - \sqrt{1-64\pi\xi(\rho+P)(e^B-1)}\right],
\label{AlBl}
\end{eqnarray}}
{and}
{
\begin{equation}
    2\, P' + (P+\rho)A' = 0,
\label{ce1}    
\end{equation}}
{the conservation equation.}

{Note that this system of equations does not explicitly depend on the metric function $A$. Consequently, one only needs to integrate equation (\ref{All}) numerically to obtain $A'(r)$, so that it effectively behaves as a first-order differential equation for $A'$. Once $A'(r)$ is known, equation (\ref{AlBl}) can be numerically integrated to obtain $B(r)$, which is also governed by a first-order differential equation.}

{An expression for the nonmetricity scalar can be easily obtained by substituting equation (\ref{AlBl}) into equation (\ref{Qr}), namely,
\begin{eqnarray}
Q(r) = \frac{1}{4\xi}\left[ \sqrt{1-64\pi\xi(\rho+P)(e^B-1)} - 1\right].
\label{QrN}
\end{eqnarray}
This expression shows that $Q(r) \leq 0$ for any value of $\xi$. Note that $Q(0)=0$, since $B(0)=0$ by regularity at the origin. Another important result is that outside the matter distribution $Q(r \ge R)=0$.}

{As already mentioned in Section \ref{be}, the vacuum solutions in $f(Q)$ gravity coincide with those of GR.}

{The vacuum solution can be explicitly obtained by setting $\rho=0$ in equation (\ref{All}), resulting in
\begin{eqnarray}
A'' = - \frac{A'}{r} - \frac{e^{-A}}{e^{-A}-1} A'^2\ .
\end{eqnarray}
A similar differential equation can be obtained for $B$. In addition, equation (\ref{AlBl}) in vacuum implies that $A' + B' = 0$. The time coordinate can then be chosen so that $A + B = 0$. Consequently, one obtains the following equation for $B''$:
\begin{eqnarray}
B'' = - \frac{B'}{r} + \frac{e^{B}}{e^{B}-1} B'^2\ .
\end{eqnarray}
As shown in detail in \cite{FA6}, the solution of these last two equations is given by
\begin{eqnarray}
e^A = e^{-B} = 1 - \frac{2 M_S}{r},
\label{MS}
\end{eqnarray}
which coincides with the Schwarzschild solution of GR. Therefore, $M_S$ is interpreted as the gravitational mass, obtained from the spacetime geometry.}

{Another useful mass definition is the total rest mass $M_0$, which is obtained by integrating the following differential equation:
\begin{eqnarray}
\frac{dm_0}{dr}=4\pi \rho_0 \,e^{B/2} r^2 \, ,
\label{dm0dr}
\end{eqnarray}
where $\rho_0$ is the rest-mass density and $4\pi e^{B/2} r^2 dr$ is the proper volume element. For bound configurations, one must have $M_S < M_0$. Thus, this condition must be satisfied for any neutron-star model.}

\section{Neutron stars with realistic equations of state}
\label{Ne}

{In a previous work \cite{FA6}, we analyzed neutron star models in the same $f(Q)$ framework using polytropic equations of state. Here, we extend that analysis by considering realistic EOSs.}

{In this section, we present numerical models of spherically symmetric neutron stars in covariant $f(Q)=Q+\xi Q^2$ gravity. We analyze the mass--radius ($M_S \times R$) and mass--central density ($M_S \times \rho_c$) relations for different values of the parameters $\xi = -2, -1, 0, 1,$ and $2$, which are given in units of the square of the gravitational radius of the Sun ($r_{gs} = 2GM_\odot/c^2 \simeq 2.95\, {\rm km})$. Note that $\xi=0$ corresponds to GR.}

{In addition, we explicitly display the behavior of the nonmetricity scalar $Q(r)$, as well as the metric functions $A(r)$ and $B(r)$, in order to provide further insight into the internal structure of the stellar configurations.}

\subsection{Realistic EOSs}
\label{realistic}

{The EOS of neutron star matter remains an open problem, particularly at extremely high densities found deep inside such stars. This uncertainty has led to the proposal of a large number of EOS models in the literature, many of which cannot yet be ruled out by current observations or experiments.}

{Since the EOS affects the predicted maximum mass, radius, and compactness of a neutron star, we consider four representative models with increasing stiffness: FPS, SLy, ENG, and MPA1.}

{The procedure used to model stars in $f(Q)$ gravity is essentially the same as in GR. The central boundary conditions $m=0$ and $P=P_c$ in $r=0$ must be specified.

In addition, central boundary conditions must be imposed for the metric functions $A'$ and $B$. Since the system now involves Eqs. (\ref{All}) and (\ref{AlBl}), regularity at the stellar center requires $A'(0)=0$ and $B(0)=0$.}

The stellar radius $R$ is defined as the value of the radial coordinate at which the pressure vanishes, i.e., $P(R)=0$, which is determined through the numerical integration of the differential equations starting from $r=0$. 

{The numerical integrations were performed using the Python \texttt{solve\_ivp} routine from the SciPy library. For each central pressure, the integration starts at the stellar center and proceeds until the pressure vanishes, determining the stellar surface and the corresponding mass and radius. We adopted the RK45 method and compared the results with those obtained using other integration methods available in \texttt{solve\_ivp}. The FPS, SLy, ENG and MPA1 equations of state were provided as tabulated data. We used cubic-spline interpolation in logarithmic variables to obtain the density values required during integration. The tables with the EOSs were obtained from the Xtreme Astrophysics neutron star database, whose EOS data are described in \cite{Ozel}.}

The gravitational mass $M  \equiv m(R)$ is obtained in the same way as in GR. However, as discussed in our previous work \cite{FA6}, the definition of mass in $f(Q)$ gravity is not unique. Since the vacuum solution is given by the Schwarzschild metric, the mass can also be obtained from the metric functions at the stellar surface, $A(R)$ or $B(R)$, leading to a geometrical definition, which we denote by $M_S$. In general, $M_S \neq M$, and we refer the reader to \cite{FA6} for a detailed discussion of this issue. 

The mass-radius and mass–central density sequences are useful tools to compare different gravitational theories with GR. In the following, we present the results obtained for each EOS.


{The sequences shown here describe equilibrium configurations but do not determine, by themselves, whether the stars are stable. In GR, the maximum-mass point along a sequence of nonrotating stars with a fixed equation of state marks the beginning of radial instability \cite{Shapiro}. However, in alternative theories of gravity, the relation between the maximum-mass point and radial stability need not be the same and must be investigated separately for each theory. Since we do not perform a stability analysis for our $f(Q)$ models, $M_{max}$ denotes only the maximum mass obtained along each equilibrium sequence. The stability of these configurations is beyond the scope of this work and will be addressed in future work.}

As already mentioned, we consider FPS, SLy, ENG, and MPA1, which have increasing stiffness. Figure \ref{EOSs} shows the GR sequences for these EOSs. As expected, stiffer EOSs lead to larger stellar radii and higher maximum masses. This specific choice is interesting for studying how the effects of the quadratic term in $Q$ are sensitive to stiffness.

{The EOSs adopted in this work satisfy the causality condition: the speed of sound in the matter inside the star remains below the speed of light, $c_s<c$. This condition is already incorporated in the construction of the tabulated EOSs considered here.}

\begin{figure}[h]
    \centering
    \begin{subfigure}[b]{0.49\textwidth}
        \centering
        \includegraphics[width=\textwidth]{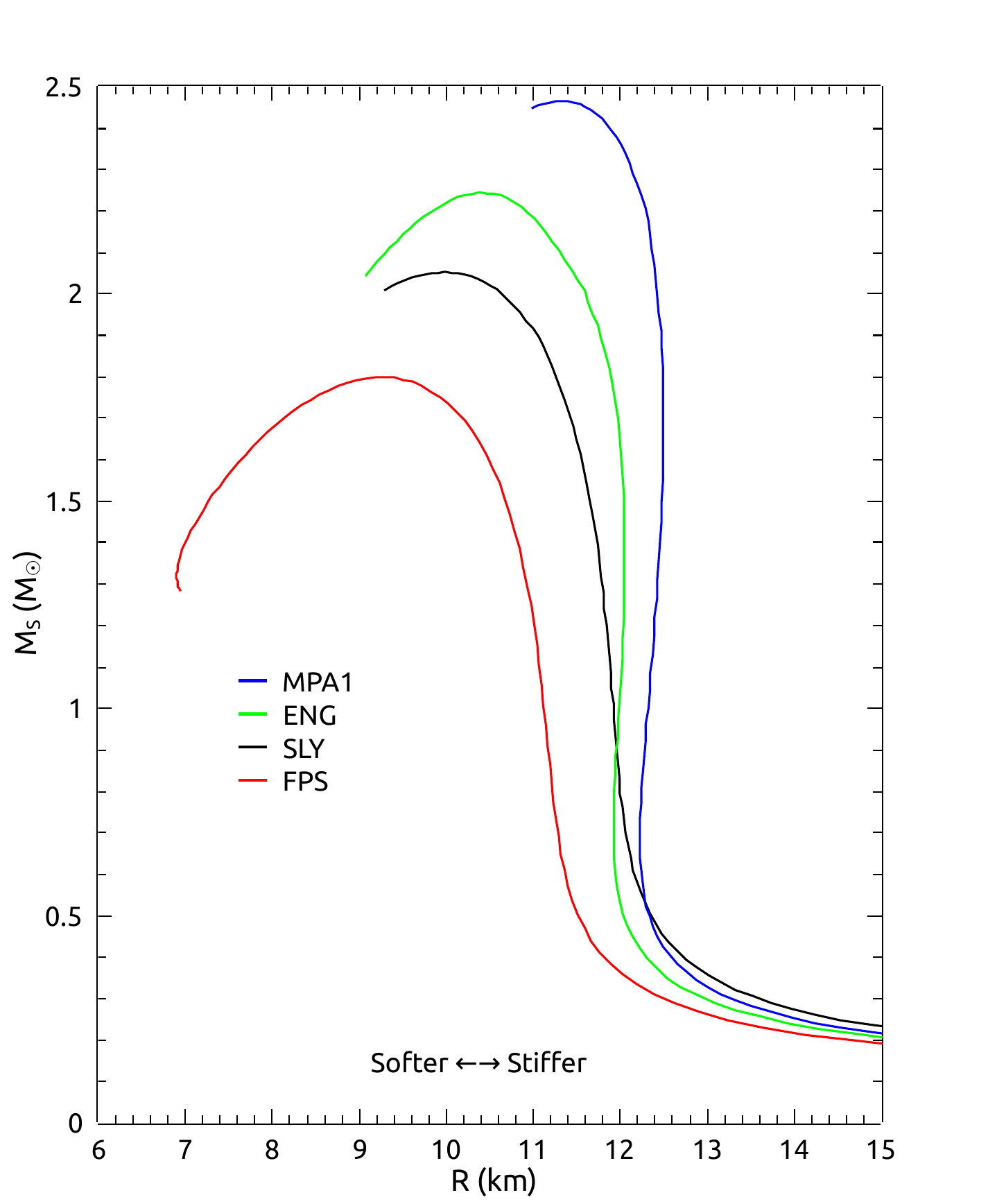}
        \label{EOSse}
    \end{subfigure}
    \hfill 
    \begin{subfigure}[b]{0.48\textwidth}
        \centering
        \includegraphics[width=\textwidth]{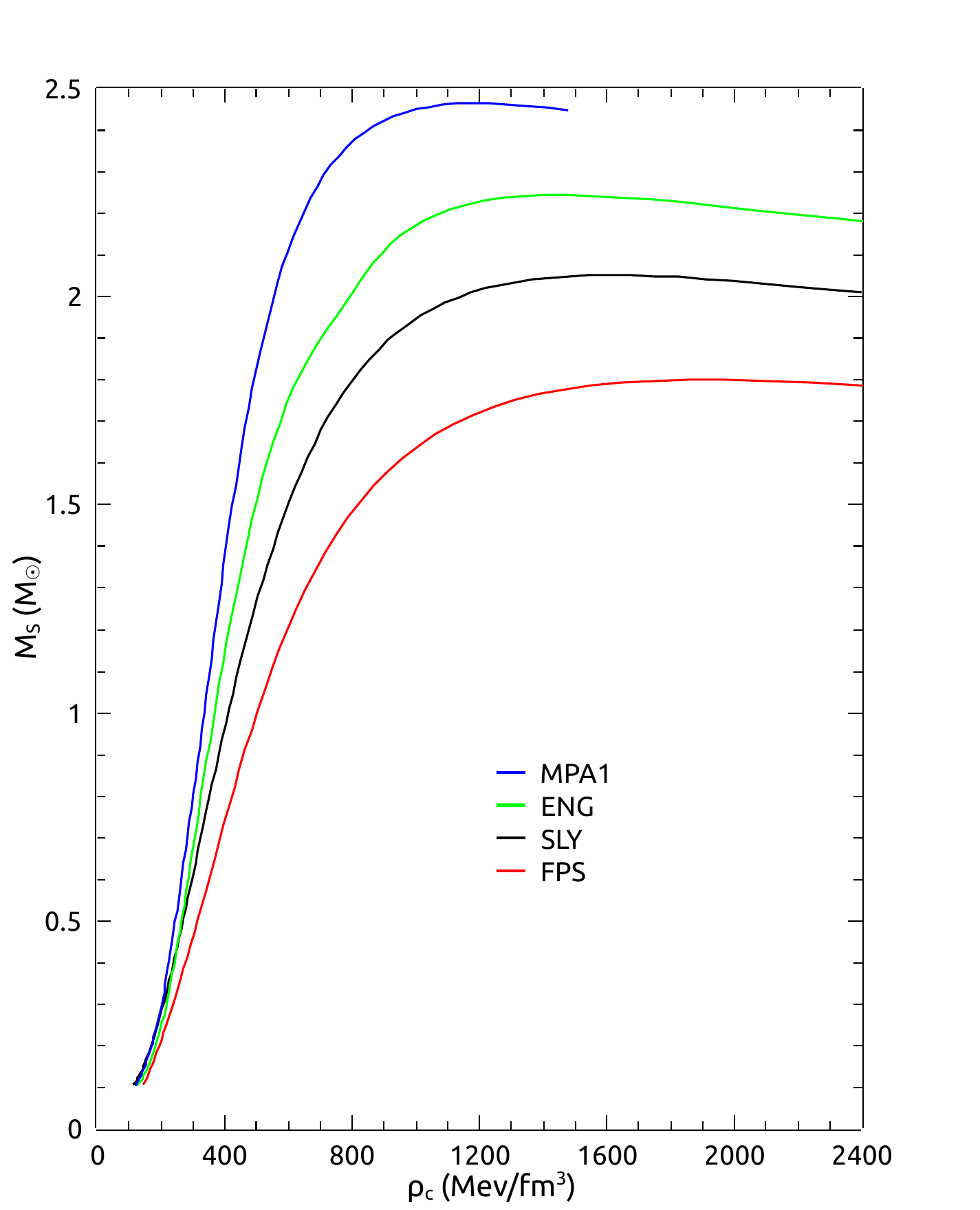}
        \label{EOSsd}
    \end{subfigure}
    \caption{Left (Right): GR sequences of M$_S$ vs. radius $R$ (central energy density $\rho_c$) for FPS, SLy, ENG, and MPA1 EOSs.}
    \label{EOSs}
\end{figure}

{We begin with the softest EOSs, namely FPS and SLy, whose results are displayed in Figures \ref{FPS} and \ref{SLY}, where we have the sequences ``Mass $\times R$'' and ``Mass $\times \rho_c$'' for different values of $\xi$, given in units of the square of the gravitational radius of the Sun ($r_{gs} = 2GM_\odot/c^2 \simeq 2.95\, {\rm km}$). In these sequences, we take the neutron star mass to be $M_S$.} 

{Notice that data from some neutron stars are also displayed in these figures, as well as in Figures \ref{ENG} and \ref{MPA1}. Later in this paper, in Subsection \ref{comparison}, we discuss a comparison between these data and our models.}

Note that we consider positive and negative values of $\xi$. Moreover, for comparison, we also consider the case in which $\xi$ vanishes, that is, GR. The mass-radius relations show a clear dependence on the parameter $\xi$. In particular, negative values of $\xi$ shift the curves towards higher masses, allowing configurations with higher maximum masses compared to GR. In contrast, positive values of $\xi$ reduce the maximum mass supported by the star. Regarding the central density, the maximum mass occurs at higher values of $\rho_c$ for $\xi < 0$.  These results indicate that in the present model, {the negative values of $\xi$ increase the maximum mass relative to the case $\xi=0$, while the positive values decrease it.

{Some configurations violate the condition $M_S < M_0$. It is well known that physically consistent configurations are expected to satisfy this condition. Therefore, the occurrence of $M_S > M_0$ may indicate that such configurations are unstable or unphysical. These regions become more significant as $\xi$ increases.}

Now we consider ENG EOS which is stiffer than FPS and SLy EOSs, whose results are shown in Figure \ref{ENG}. The same qualitative aspects observed for the softer EOSs persist in this case. Negative values of $\xi$ lead to an increase in maximum mass, while positive values reduce it. However, deviations from GR become more pronounced for the ENG EOS.

{In the mass-radius sequences, the curves corresponding to $\xi < 0$ are clearly shifted towards higher masses, with a greater increase in maximum mass compared to the FPS and SLy cases. This indicates that the effects of the quadratic correction in $Q$ become more significant in the presence of stiffer matter.}

{As before, the condition $M_S < M_0$ is not always satisfied, so the regions where $M_S > M_0$ become more evident for positive values of $\xi$.}

{Finally, we analyze the results for MPA1 EOS, presented in Figure \ref{MPA1}. This is the stiffest EOS considered in this work. The mass-radius relations exhibit the same general behavior as observed for the other EOSs. Negative values of $\xi$ lead to an increase in maximum mass, while positive values reduce it. Due to its stiffness, the maximum mass is larger than for the other EOSs. The corresponding configurations also tend to have larger radii.}

{From the sequence of EOSs considered, from FPS to MPA1, the quadratic term in $Q$ affects the stellar structure.  This effect becomes particularly relevant in the high-density regime, where the maximum mass is achieved. In particular, negative values of $\xi$ favor more massive configurations, while positive values reduce the maximum mass. The behavior of the mass-central density relations suggests that the stability properties of the configurations are also affected.}

{Table \ref{Tab} summarizes additional results for the FPS, SLy, ENG, and MPA1 EOSs. For each value of $\xi$, we present the maximum mass considered and the corresponding radius, compactness, and central density.}

{For $\xi = 0$ (GR), the maximum mass for each EOS is given by the maximum of the corresponding sequence shown in Figures \ref{FPS}, \ref{SLY}, \ref{ENG} and \ref{MPA1}.} 

{For $\xi \neq 0 $, we first check if $M_S < M_0$. Our calculations show that for the positive values of $\xi$ considered, this inequality is not always satisfied. In these cases, the maximum masses occur when $M_S = M_0$.} 

{For the negative values of $\xi$ considered here, our calculations yield $M_S < M_0$. In these cases, we adopt the maximum of the sequences as the maximum masses. Note also that for the negative values of $\xi$ considered, there is no turning point in the sequences for the SLy and MPA1 EOSs. In such cases, the maximum central densities coincide with the maximum density of the EOSs themselves.} 
      
{For negative $\xi$, the compactnesses are not significantly modified. This is because both their masses and radii increase as $\xi$ becomes more negative. This occurs in such a way that the compactnesses remain essentially unchanged.}

\vskip 12 pt

\begin{figure}[h]
    \centering
    \begin{subfigure}[b]{0.49\textwidth}
        \centering
        \includegraphics[width=\textwidth]{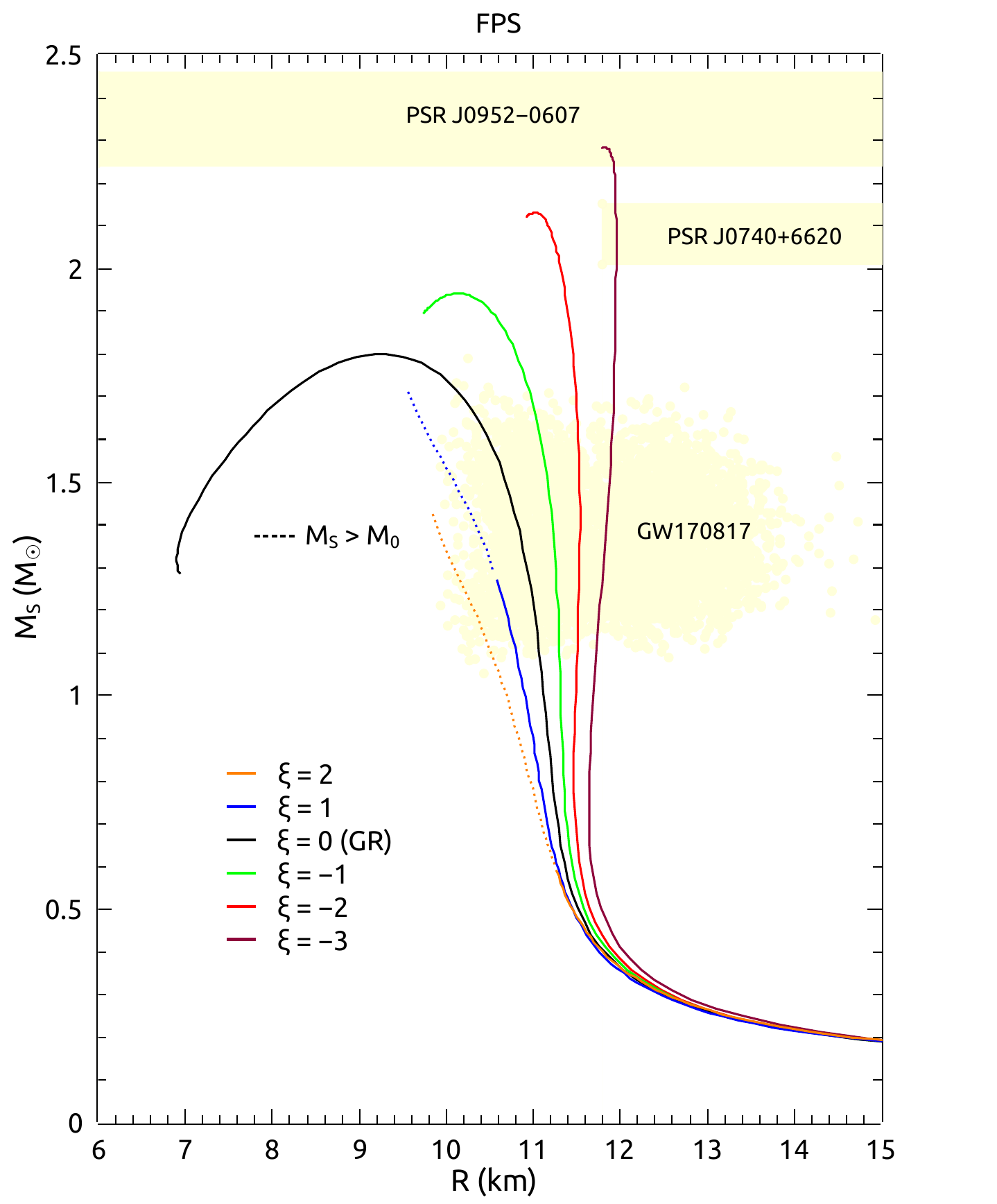}
        \label{FPSe}
    \end{subfigure}
    \hfill 
    \begin{subfigure}[b]{0.485\textwidth}
        \centering
        \includegraphics[width=\textwidth]{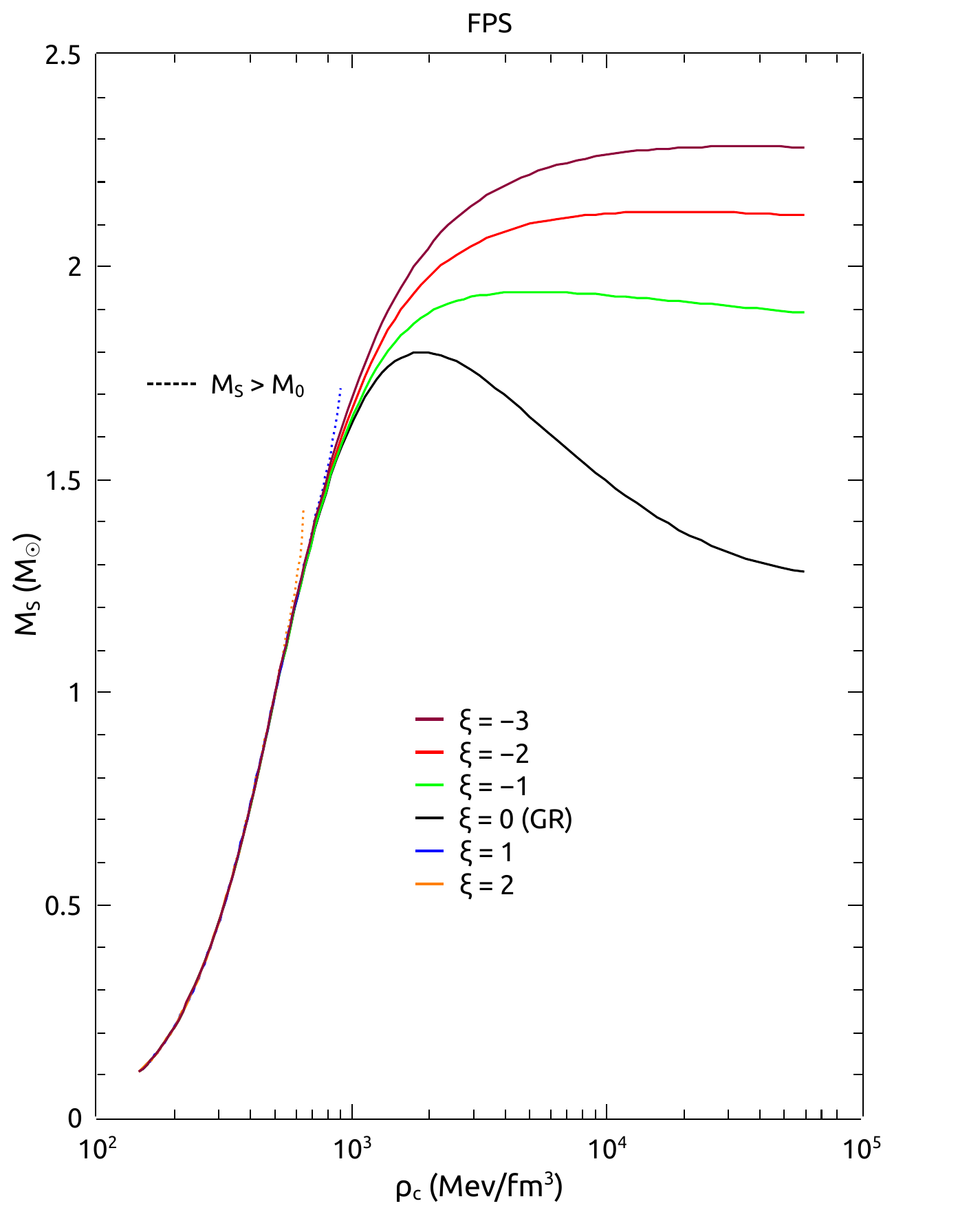}
        \label{FPSd}
    \end{subfigure}
    \caption{Left (Right): Sequences of M$_S$ vs. radius $R$ (central energy density $\rho_c$) for the FPS EOS, and different values of $\xi$, which is given in units of the square of the gravitational radius of the Sun ($r_{gs} = 2GM_\odot/c^2 \simeq 2.95\, {\rm km})$.}
    \label{FPS}
\end{figure}

\begin{figure}[h]
    \centering
    \begin{subfigure}[b]{0.49\textwidth}
        \centering
        \includegraphics[width=\textwidth]{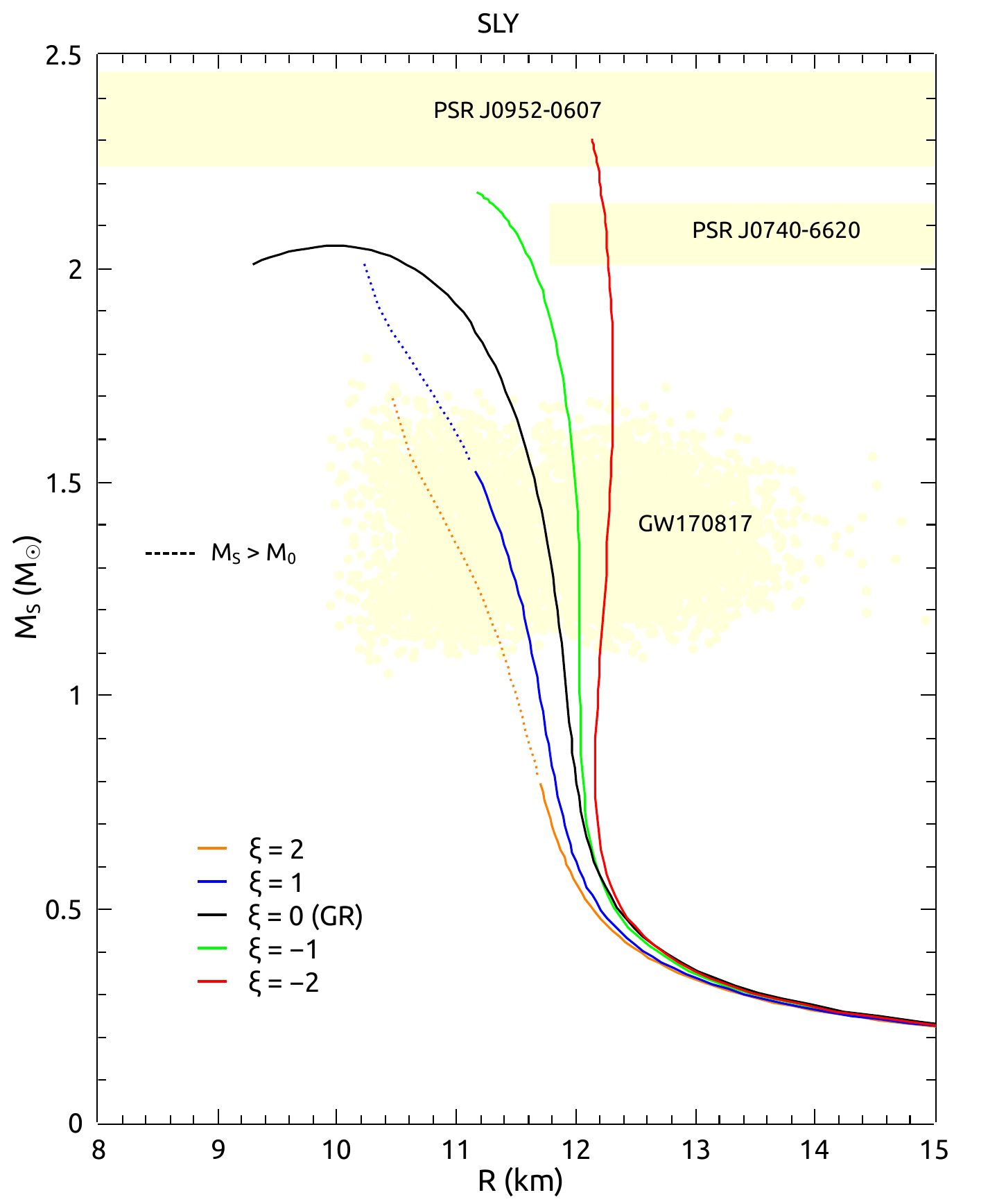}
        \label{SLYe}
    \end{subfigure}
    \hfill 
    \begin{subfigure}[b]{0.485\textwidth}
        \centering
        \includegraphics[width=\textwidth]{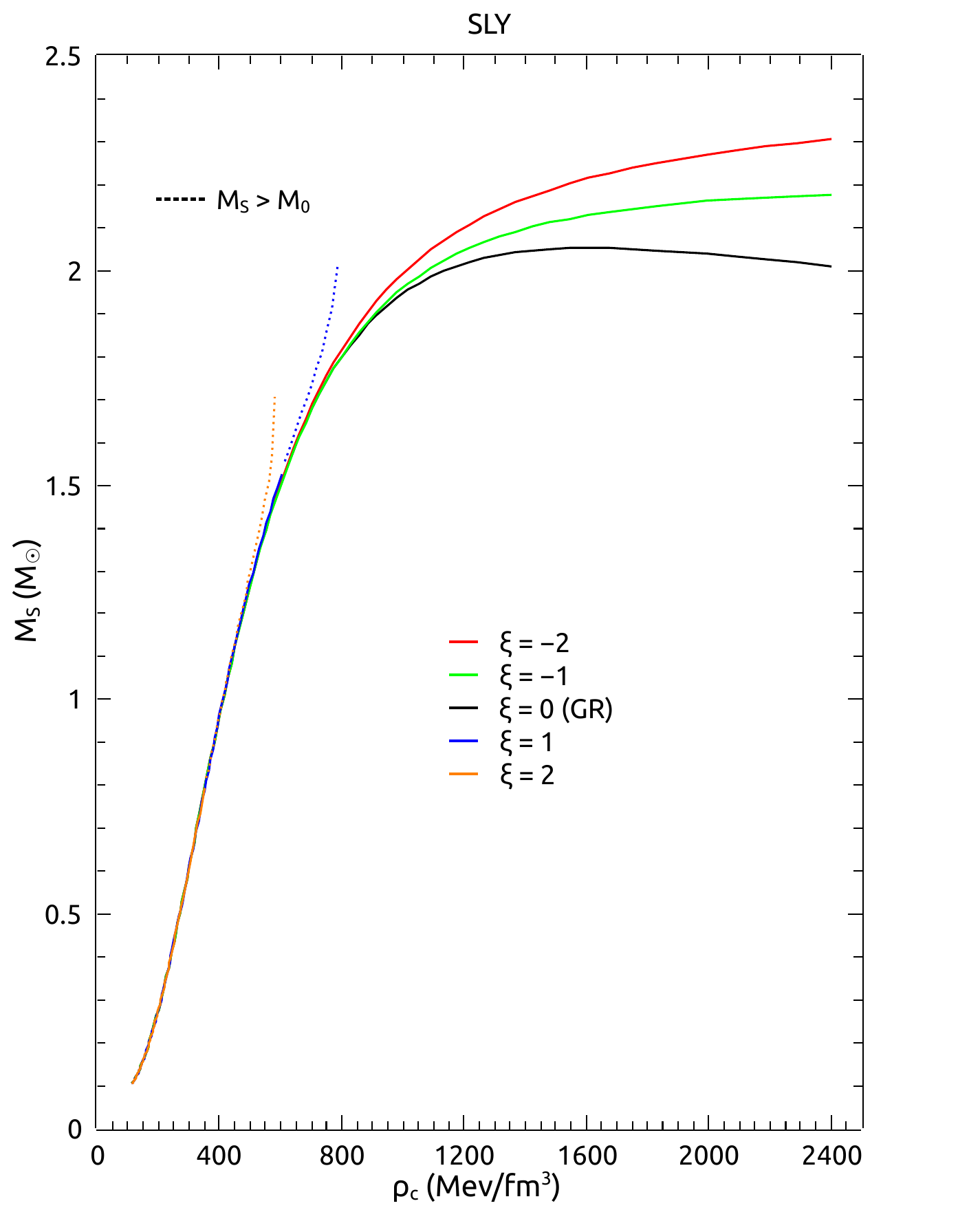}
        \label{SLYd}
    \end{subfigure}
    \caption{The same as in Figure \ref{FPS} for SLy EOS.}
    \label{SLY}
\end{figure}

\begin{figure}[h]
    \centering
    \begin{subfigure}[b]{0.49\textwidth}
        \centering
        \includegraphics[width=\textwidth]{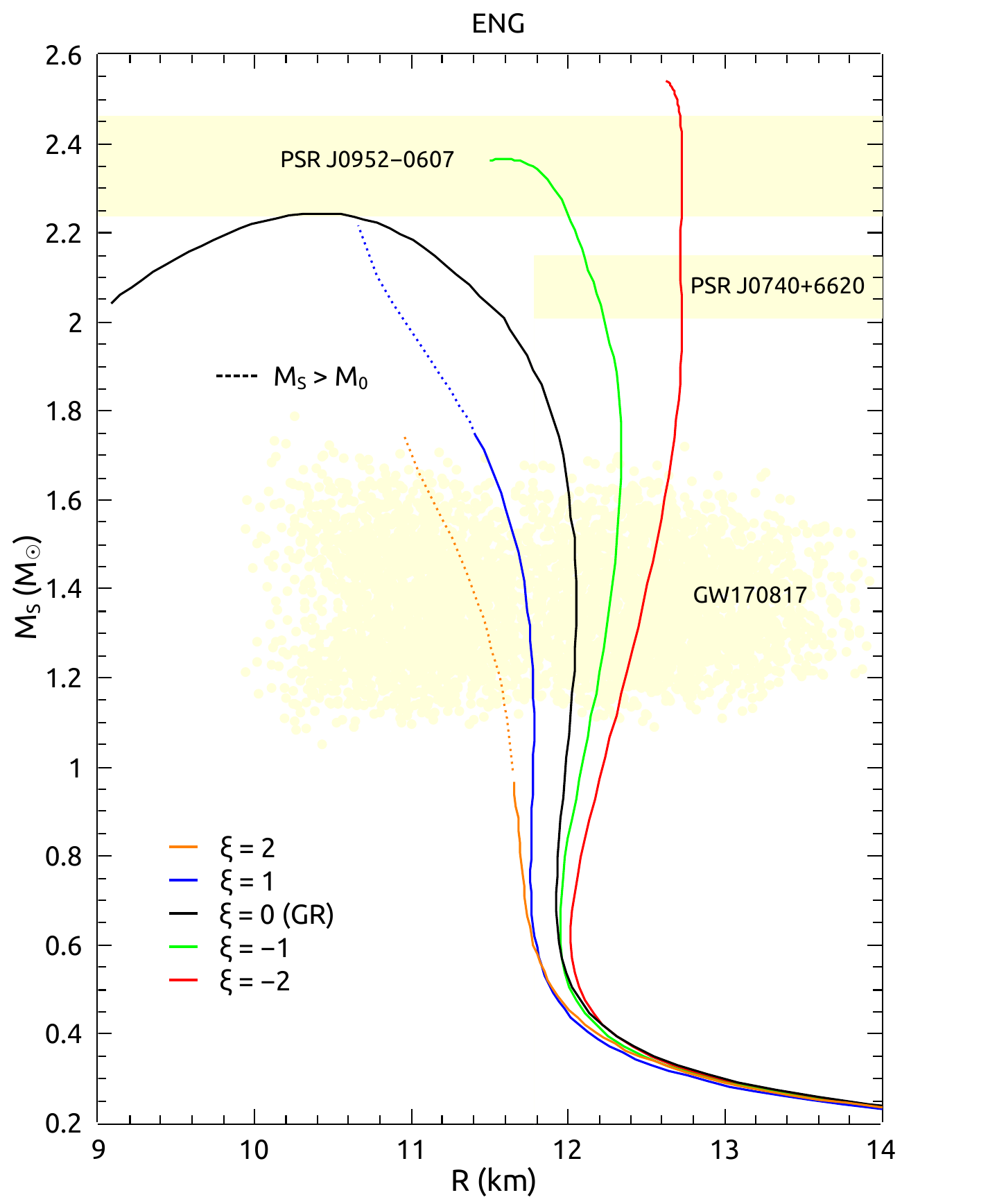}
        \label{EMGMR}
    \end{subfigure}
    \hfill 
    \begin{subfigure}[b]{0.48\textwidth}
        \centering
        \includegraphics[width=\textwidth]{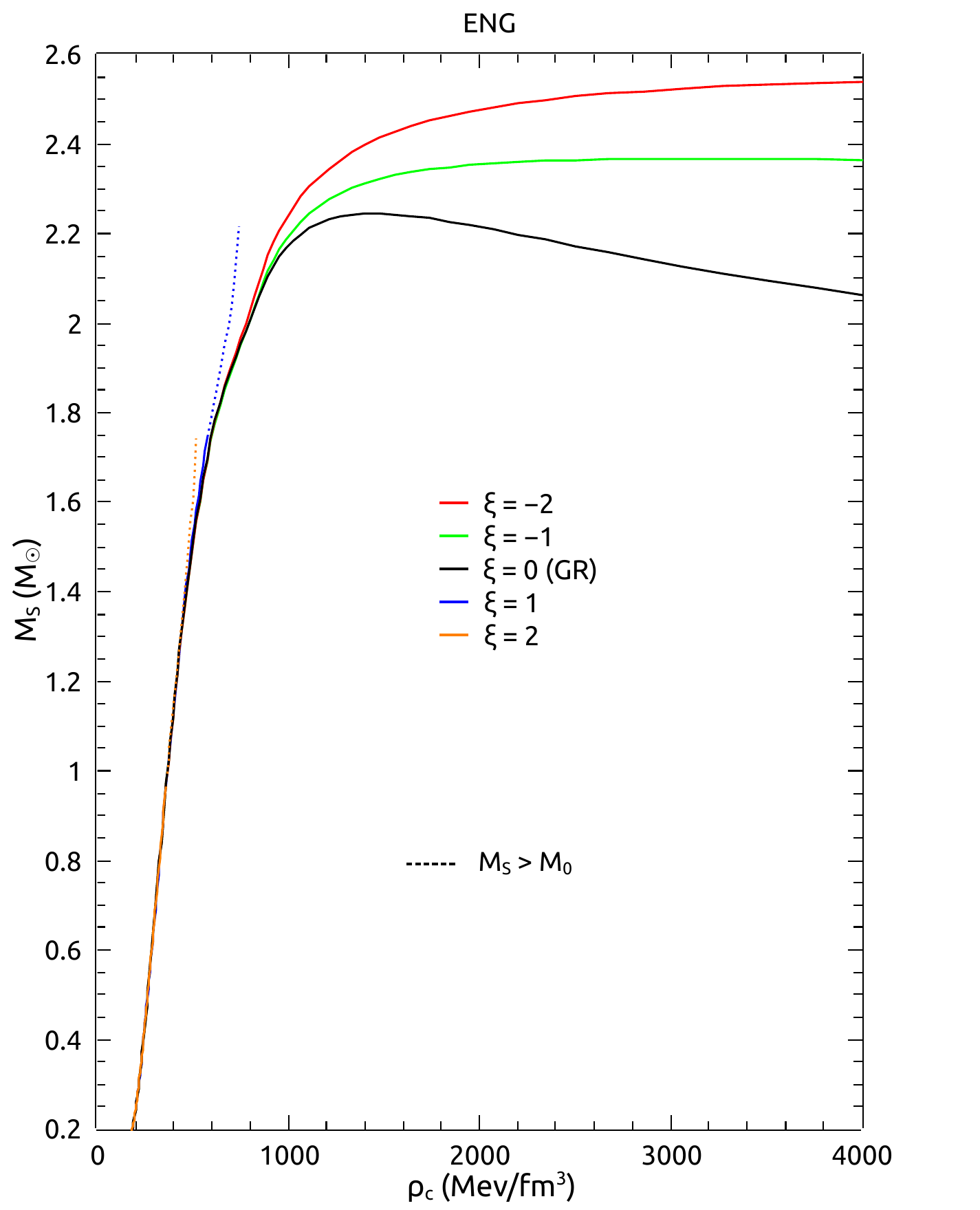}
        \label{ENGd}
    \end{subfigure}
    \caption{The same as in Figure \ref{FPS} for ENG EOS.}
    \label{ENG}
\end{figure}

\begin{figure}[h]
    \centering
    \begin{subfigure}[b]{0.49\textwidth}
        \centering
        \includegraphics[width=\textwidth]{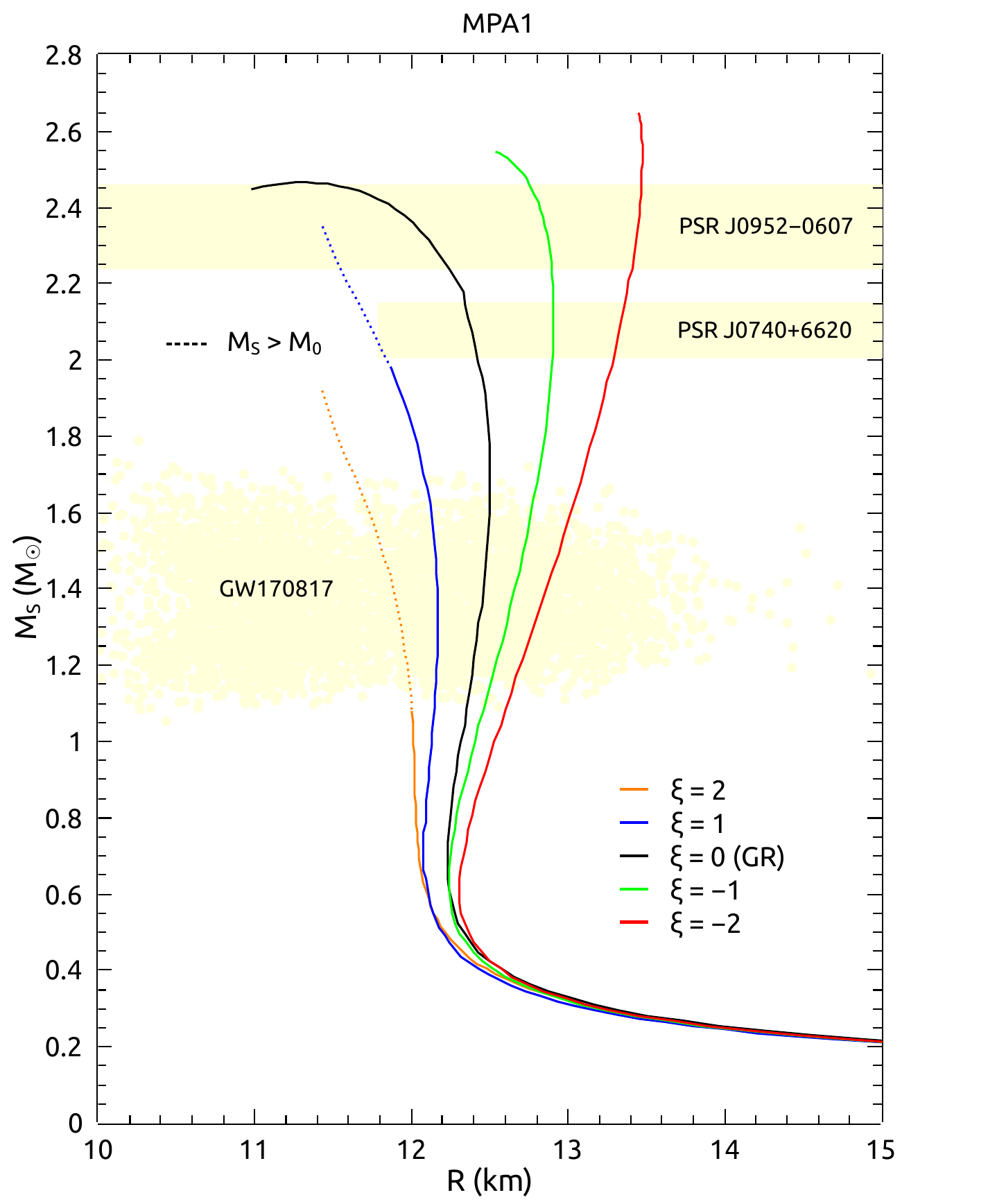}
        \label{MPA1MR}
    \end{subfigure}
    \hfill 
    \begin{subfigure}[b]{0.485\textwidth}
        \centering
        \includegraphics[width=\textwidth]{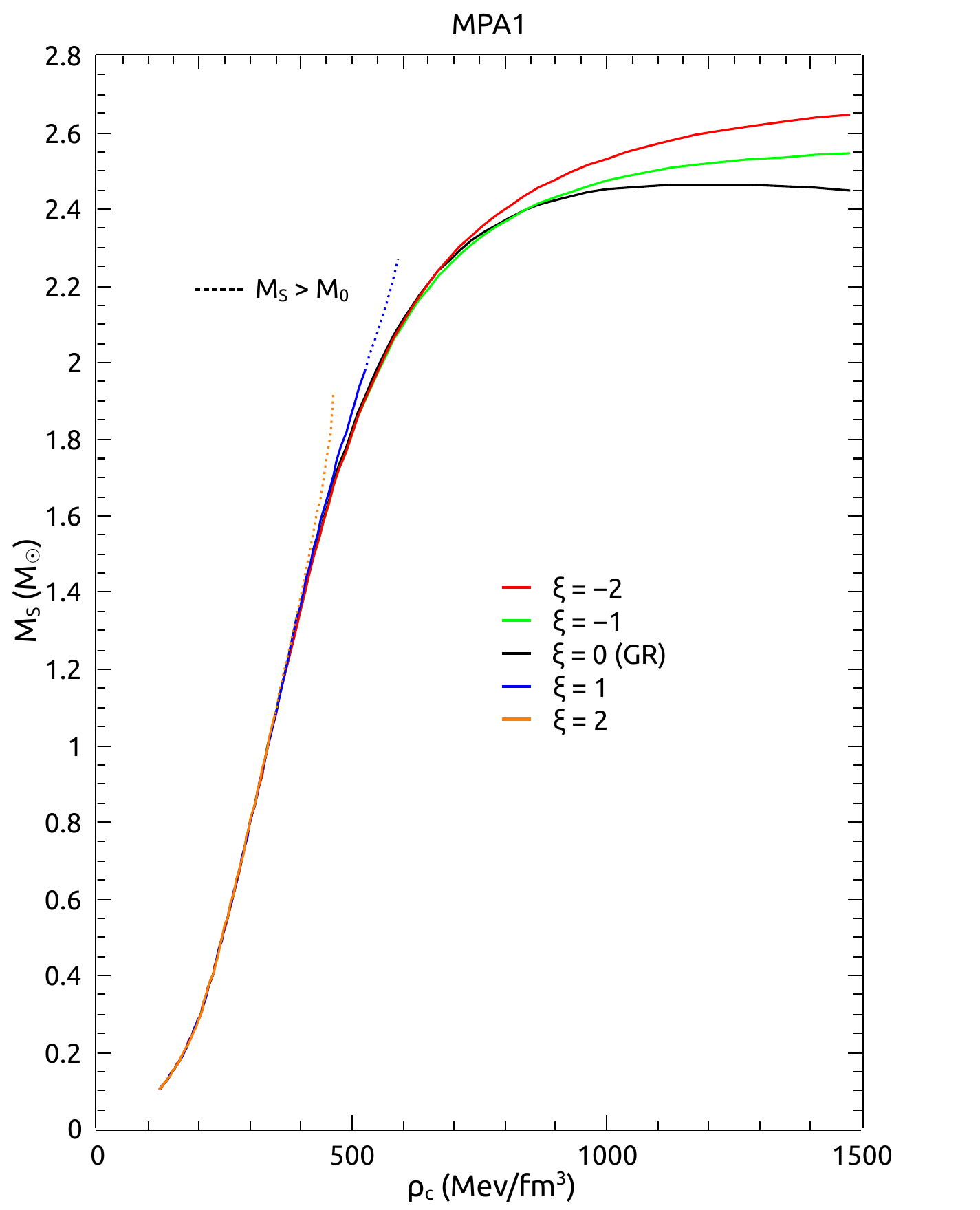}
        \label{MPA1d}
    \end{subfigure}
    \caption{The same as in Figure \ref{FPS} for MPA1 EOS.}
    \label{MPA1}
\end{figure}

\begin{table}[]
    \centering
        \caption{Maximum masses and their corresponding radii, dimensionless compactnesses and central densities for different values of $\xi$ for FPS, SLY, ENG, and MPA1 EOSs.}
    \begin{tabular}{c c c c c}
    \hline \hline
   $\xi$ ($r_{gs}^2$) & $M_{max}/M_\odot$ & R (km) & M/R  & $\rho_c$ (GeV/fm$^3$) \\ 
     \hline \hline
    & & FPS & & \\   
    \hline  \hline
      -3 & 2.28  & 11.81  & 0.29 & 43 \\     
      -2 & 2.13  & 11.02  & 0.29 & 19 \\
      -1 & 1.94  & 10.12  & 0.28 & 5.4 \\
       0 & 1.80  & 9.30  & 0.29 & 1.9 \\
       1 & 1.27  & 10.58  & 0.18 & 0.63 \\
       2 & 0.59  & 11.26  & 0.08 & 0.35 \\
       \hline  \hline
      & & SLY & & \\    
       \hline \hline
      -2 & 2.31  & 12.14  & 0.28 & 2.4  \\
      -1 & 2.18  & 11.17  & 0.29 & 2.4  \\
       0 & 2.05  & 9.98   & 0.30 & 1.6  \\
       1 & 1.53  & 11.16  & 0.20 & 0.60 \\
       2 & 0.80  & 11.70  & 0.10 & 0.35 \\
       \hline  \hline
      & & ENG & & \\    
       \hline \hline
      -2 & 2.54  & 12.63  & 0.30 & 4.4  \\
      -1 & 2.37  & 11.60  & 0.30 & 3.1  \\
       0 & 2.24  & 10.39  & 0.31 & 1.5  \\
       1 & 1.75  & 11.41  & 0.23 & 0.58 \\
       2 & 0.97  & 11.66  & 0.12 & 0.36 \\
       \hline  \hline
      & & MPA1 & & \\    
       \hline \hline
       -2 &  2.65  & 13.45  & 0.29  & 1.5  \\
      -1 &  2.55  & 12.54  & 0.30  & 1.5  \\
       0 &  2.47  & 11.33  & 0.32  & 1.2  \\
       1 &  1.98  & 11.87  & 0.25  & 0.52 \\
       2 &  1.08  & 12.00  & 0.13  & 0.35 \\
       \hline \hline
    \end{tabular}
\label{Tab}
\end{table}

\subsection{Comparison with neutron star observations}
\label{comparison}

{Neutron stars with masses close to or above $2$ M$_\odot$ provide important constraints on the EOS of dense matter. Within GR, only extremely stiff EOSs such as MPA1 can explain such massive neutron stars (see, e.g., Figure \ref{EOSs}). However, these massive stars can be explained, even for soft EOSs, in modified gravity.}

{From gravitational wave observations, there is evidence that in some of the detected events, some of the components of binary systems are neutron stars. In the GW170817 \cite{GW170817}, GW190425 \cite{GW190425} and GW190814 \cite{GW} events, for example, neutron stars may be present. In GW170817, both components of the associated binary system are consistent with neutron stars of masses around 1.4 M$_\odot$. However, the GW190425 and GW190814 events may be associated with neutron stars with masses around 2.5 M$_\odot$ or greater.}

{Several pulsars have masses estimated from observational data. In particular, PSR J1614-2230 \cite{Demorest}, PSR J0740+6620 \cite{Cromartie}, PSR J0348+0432 \cite{Antoniadis} and PSR J0952-0607 \cite{Romani} have inferred masses near or above $2$ M$_\odot$. Recall that PSR J0952-0607 is the fastest (with a rotation period of 1.41 ms) and heaviest (2.35 $\pm$ 0.11 M$_\odot$) known Galactic neutron star. Consequently, these pulsars challenge both EOS models and GR.}

{With the advent of NICER (Neutron Star Interior Composition Explorer), a NASA X-ray telescope on board the ISS (International Space Station), it is possible to obtain the radii of a couple of neutron stars (see, e.g., \cite{Miller}). In particular, Ref. \cite{Miller} summarizes the results for PSR J0030+0451 \cite{Kini}, PSR J0740+6620 \cite{Dittmann}, PSR J0437-4715 \cite{PSRJ0437}, PSR J1231-1411 \cite{PSRJ1231}, and PSR J0614-3329 \cite{Mauviard}.}

{Instead of including all the data for the neutron stars mentioned above in Figures \ref{FPS}, \ref{SLY}, \ref{ENG}, and \ref{MPA1}, we included some representative data. As can be seen, we included data for PSR J0952-0607, which also represent GW190425 and GW190814, whose estimated masses are similar. With the exception of PSR J0740+6620, all other pulsars studied by NICER can be represented by GW170817\footnote{The posterior of GW170817 can be found at \url{https://dcc.ligo.org/public/0152/P1800115/005/Parametrized-EoS_maxmass_posterior_samples.dat}.}, since their masses and radii are approximately within the posterior distribution of this GW event. In addition, we included data for PSR J0740+6620, which has an estimated mass $2.08 \pm 0.07$ M$_\odot$ and a radius between 12 and 15 km \cite{Miller, Dittmann}.}

{As is well known, GR cannot explain the most massive neutron stars with soft EOSs (see, e.g., Figures \ref{FPS} and \ref{SLY}). Among the EOSs considered, only the stiff MPA1 EOS supports such high masses in GR (see Figure \ref{MPA1}). On the other hand, the $f(Q)$ models considered here can support stars with similarly high masses even with softer EOSs.}

{Our results also show that, for softer EOSs, more negative values of $\xi$ are required to reach the masses of the most massive neutron stars.}

{It should be noted that the tidal deformability inferred from the GW170817 event provides constraints on the EOS \cite{GW170817}. However, since we are dealing with $f(Q)$ gravity, it is necessary to study the tidal deformation in this gravity, which is beyond the scope of the present work and is left for future study.}
\vskip 12pt

\subsection{Nonmetricity}

{It is useful to analyze, for the realistic equations of state considered in this work, the behavior of the nonmetricity scalar $Q(r)$, as well as the metric functions $A(r)$ and $B(r)$, for different values of the parameter $\xi$.}

{From Eq. (\ref{QrN}), one has $Q(r) \leq 0$ throughout the stellar interior, while $Q(r) = 0$ for $r \geq R$. This behavior reflects the fact that the exterior solution reduces to the Schwarzschild spacetime. Consequently, deviations from GR are confined to the interior of the star and therefore affect the mass-radius relation}.

{In Fig. \ref{Q_r}, we show the radial profile of $Q(r)$ for neutron star configurations corresponding to the maximum mass $M_S$, for different values of $\xi$ and for the SLy EOS. The nonmetricity scalar exhibits a profile resembling a potential well: it vanishes at the center, reaches a minimum inside the star and then increases monotonically towards the surface, where it vanishes smoothly. Note that the magnitude of $Q(r)$ is maximal in an intermediate region inside the star. This indicates that the deviations from GR are not solely determined by the central density but instead result from the combined effect of the matter distribution and the metric functions. These effects become negligible in the outer layers, where $Q(r)$ approaches zero. {Moreover, for the EOSs and values of $\xi$ considered, the maximum mass increases as $\xi$ becomes more negative and decreases for positive $\xi$, relative to the case $\xi=0$ (GR).}

{It is interesting to compare this behavior with the corresponding results obtained for polytropic equations of state in our previous work \cite{FA6}. In that case, the minimum of $Q(r)$ occurs at smaller radii, closer to the stellar center. In contrast, for the realistic equations of state considered here, the minimum is shifted towards larger radii, indicating that the region where deviations from GR are most pronounced depends sensitively on the internal structure of the star. 

{Regarding the metric functions, although the system of equations does not explicitly depend on $A(r)$, this function can be obtained once $A'(r)$ is known from numerical integration. As in GR, $A(r)$ is defined up to a constant, which is fixed by matching the interior solution to the Schwarzschild exterior on the stellar surface. On the other hand, the metric function $B(r)$ is fully determined by integrating Eq. (\ref{AlBl}) together with the regularity condition $B(0) = 0$. The use of this first-order equation simplifies the numerical procedure and ensures a smooth behavior of the solution throughout the stellar interior.}

{In Fig. \ref{ab_r}, we present the profiles of $A(r)$ and $B(r)$ for the same configurations considered in Fig. \ref{Q_r}. Both functions are continuous and smooth on the stellar surface, with no discontinuities in their first derivatives.}

{The comparison between different values of $\xi$ reveals deviations from GR. In particular, for $\xi > 0$, the function $A(r)$ lies above the corresponding GR curve, while for $\xi < 0$ it lies below it. In contrast, the function $B(r)$ exhibits the opposite behavior.}

\begin{figure}
\centering
\includegraphics[scale=0.3]{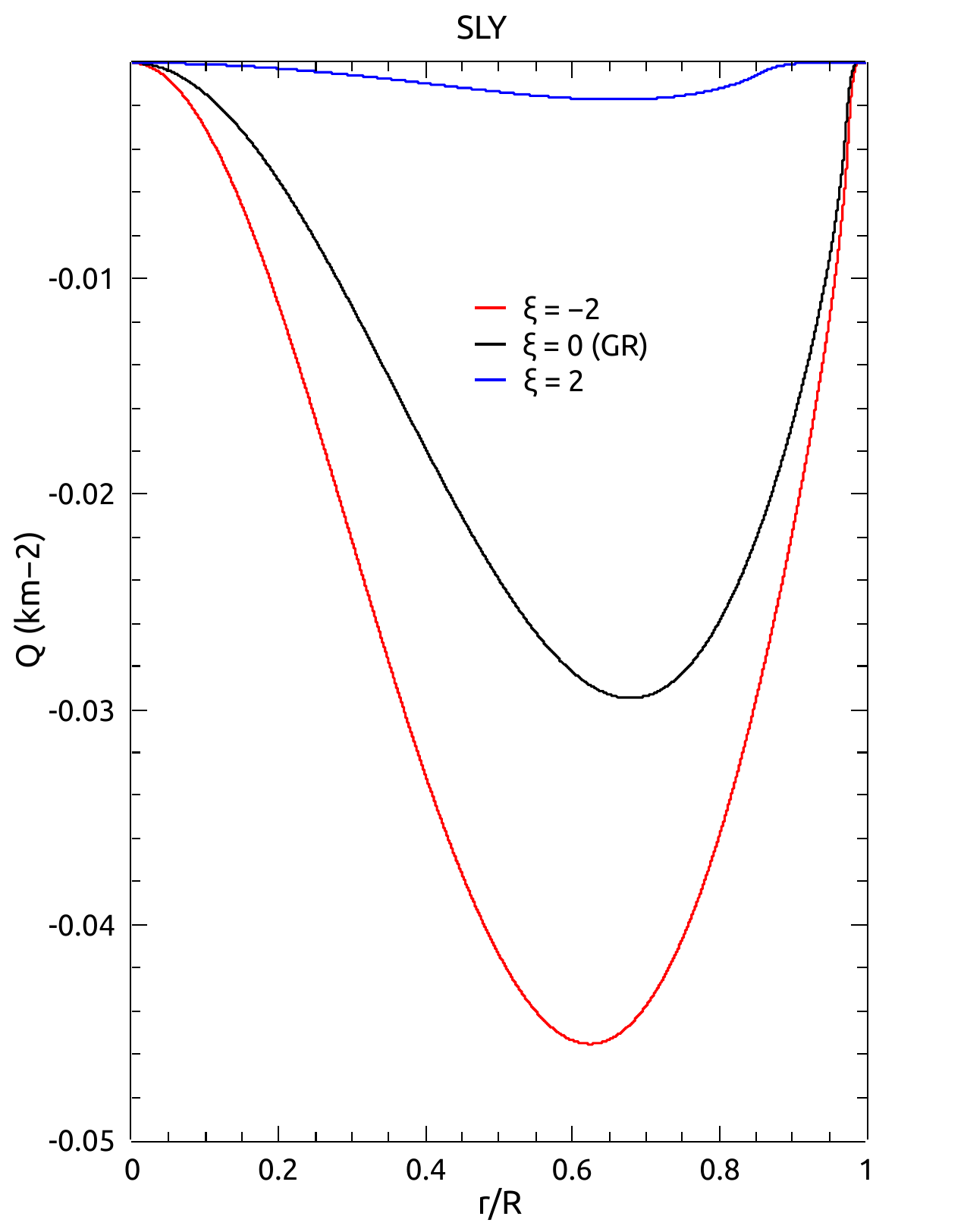}
\caption{Q(r) for maximum $M_S$ of $\xi = - 2, 0$ and $2$ for SLy EOS.}
    \label{Q_r}
\end{figure}

\begin{figure}[h]
    \centering
    \begin{subfigure}[b]{0.49\textwidth}
        \centering
        \includegraphics[width=\textwidth]{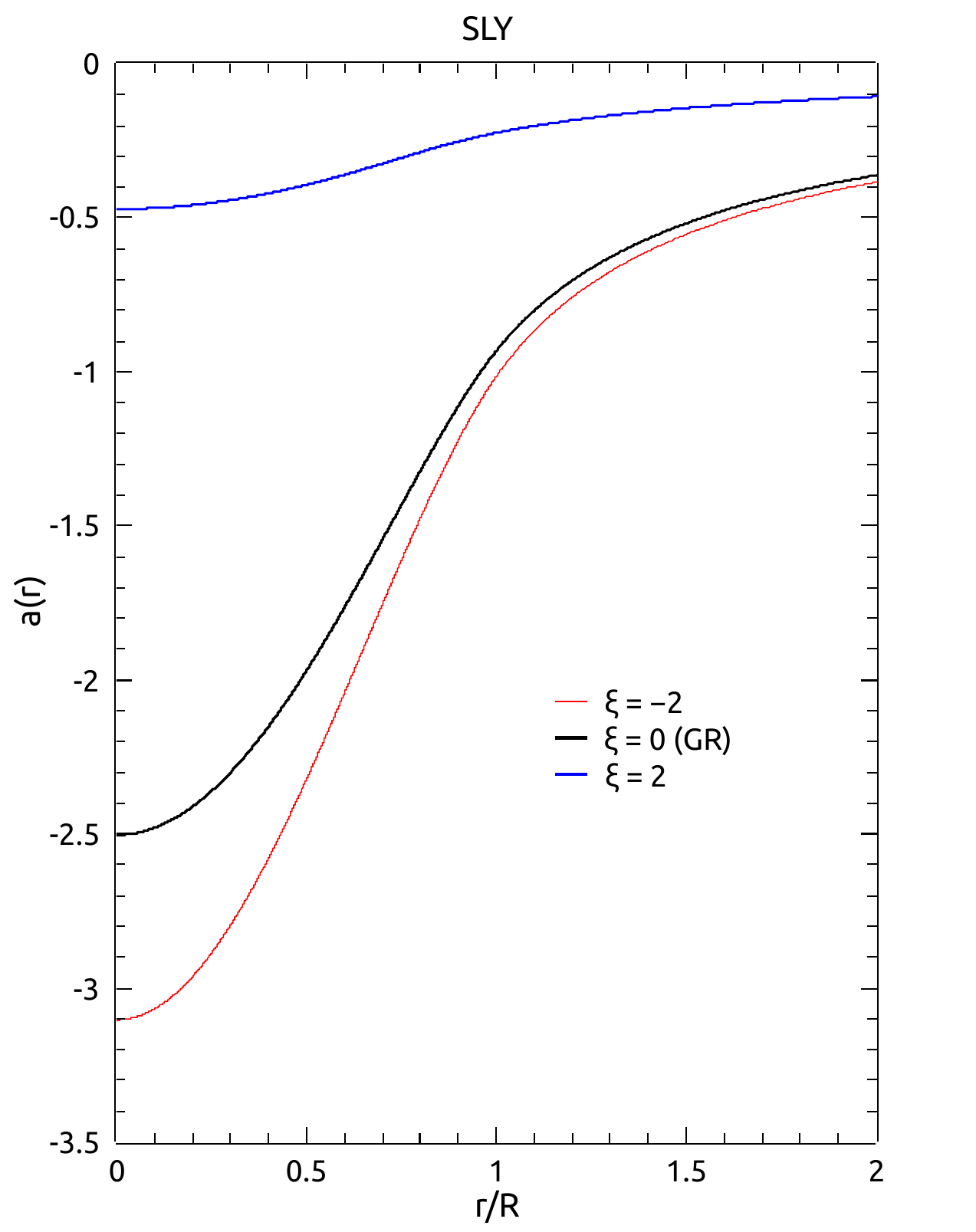}
        \label{a_r}
    \end{subfigure}
    \hfill 
    \begin{subfigure}[b]{0.47\textwidth}
        \centering
        \includegraphics[width=\textwidth]{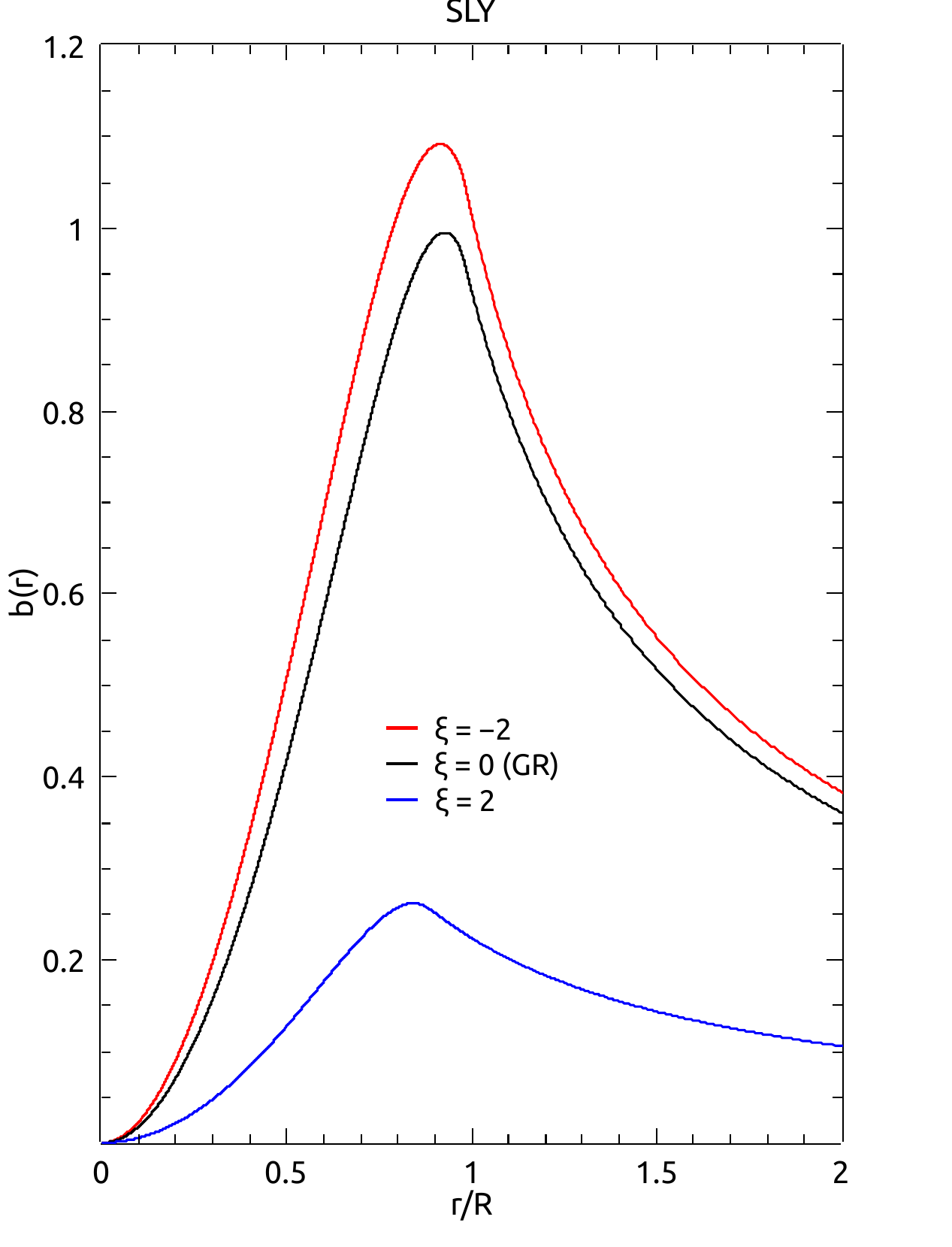}
        \label{b_r}
    \end{subfigure}
    \caption{$A(r)$ and $B(r)$ for maximum $M_S$ of $\xi = - 2, 0$ and $2$ for SLy EOS. }
    \label{ab_r}
\end{figure}

\section{Final remarks}
\label{Fr}

{Despite its success, GR does not provide a complete explanation for several observational and theoretical issues. This motivates the investigation of alternative theories of gravity. One of its limitations is related to the explanation of very massive neutron stars. Several pulsars have measured masses of about $ 2$ M$_\odot$ or higher. Pulsars such as PSR J1614-2230 \cite{Demorest}, PSR J0740+6620 \cite{Cromartie}, PSR J0348+0432 \cite{Antoniadis}, and PSR J0952-0607 \cite{Romani} are examples of such massive compact objects. Furthermore, if systems such as GW170817 \cite{GW170817}, GW190425 \cite{GW190425}, and GW190814 \cite{GW} are composed of neutron stars with masses of nearly three solar masses or result in such stars, GR does face difficulties in explaining them.}

{In this work, we have investigated neutron star configurations in the framework of $f(Q)$ gravity, considering the quadratic model $f(Q) = Q + \xi Q^2$, using realistic EOSs. The analysis extends our previous results obtained with polytropic EOSs, allowing for a more accurate description of the internal structure of compact stars.}

{Our results show that the quadratic correction to the nonmetricity scalar has important implications for the equilibrium configuration of the star. In particular, negative values of $\xi$ increase the maximum masses allowed by all realistic equations of state considered, while positive values produce the opposite effect.}

{Another general conclusion is that increasingly negative values of $\xi$ produce an effect equivalent to ``stiffening'' the EOS, leading to larger maximum masses.}

{We have also analyzed the behavior of the nonmetricity scalar and the metric functions throughout the stellar interior. The nonmetricity vanishes both at the stellar center and outside the matter distribution, reaching its maximum absolute value inside the star. This indicates that deviations from GR are confined to the high-density interior of the star and naturally disappear in vacuum, where the Schwarzschild solution is recovered.}

{Our results indicate that quadratic $f(Q)$ gravity provides a viable framework for describing neutron stars while allowing significant deviations from GR in the strong-field regime. Since the effects of the quadratic correction become relevant only at high densities, compact stars are an excellent laboratory for testing modified gravity theories.}

\section*{Data availability statement}
In this study, no new data was created or analyzed.
 
\section*{Acknowledgment}
J.C.N.A. thanks CNPq (307803/2022-8 and 308178/2026-2) for partial financial support. {We thank the referee for the detailed and constructive comments, which helped improve the manuscript.}

\end{document}